# Phonon Confinement in Stressed Silicon Nanocluster

Satyaprakash Sahoo*, S. Dhara, S. Mahadevan[1] and A. K. Arora

Materials Science Division, Indira Gandhi Centre for Atomic Research,

Kalpakkam 603102, India

[1]Nondestructive Evaluation Division, Indira Gandhi Centre for Atomic Research,

Kalpakkam 603102, India

**Abstract**

Confined acoustic and optical phonons in Si nanoclusters embedded in sapphire, synthesized using ion-beam implantation are investigated using Raman spectroscopy. The $l = 0$ and $l = 2$ confined acoustic phonons, found at low Raman shift, are analyzed using complex frequency model and the size of the nanoparticles are estimated as 4 and 6 nm. For the confined optical phonon, in contrast to expected red shift, the Raman line shape shows a substantial blue shift, which is attributed to size dependent compressive stress in the nanoparticles. The calculated Raman line shape for the stressed nanoparticles fits well to data. The sizes of Si nanoparticles obtained using complex frequency model are consistent with the size estimated from the fitting of confined optical phonon line shapes and those found from X-ray diffraction and TEM.

Keywords: Raman spectroscopy, Si nanocluster, phonon confinement, ion implantation

*Corresponding author: Satyaprakash Sahoo

E-mail: satyasahoo@igcar.gov.in, satya504@gmail.com

## I. INTRODUCTION

Nanocrystalline (nc-) Si is of considerable current interest because of its potential applications in future high-speed micro-electronic device. When the size of Si reduces in the scale to few nm, its opto-electronic properties change dramatically over its bulk value. Apart from the well known size-induced quantum confinement in nc-Si,[1-3] there are several studies on the vibrational properties of nc-Si.[4-7] In several cases asymmetrically broadened and red-shift Raman spectrum, characteristic of a phonon confinement has been found. Paillard[4] showed for size-selected cluster-assembled Si-nanoparticles that the particle size could be accurately estimated from the red-shift of the Raman peak position. In another study on Si-doped $SiO_2$ films, a comparison of Raman spectra with those of theoretically calculated vibrational density of state of $Si_{33}$ and $Si_{45}$ suggested the presence of such clusters in the film.[8] On the other hand, Si-nano-beads embedded in $SiO_2$ nanowires of varying diameter exhibited a blue shift of Raman peak, which is attributed to existence of stress in the Si-nanocrystals.[9] However, the cause of the varying magnitude of stress, i.e. whether it arises from different bead diameter or from different shell thickness around the bead, is not well understood. Furthermore, confined acoustic phonons in nc-Si have not been studied. In order to resolve the ambiguity about the origin of stress, we have investigated Si-nanocrystals of different sizes buried deep into a crystalline host. Strained nc-Si is of interest also from application point of view as the transport properties such as carrier mobility and piezo-resistance shows improved characteristic.[10-12] Stress in nc-Si are also known to result in high-pressure phase of Si.

In the present work we report a study of Si-nanoclusters embedded in a single crysral of $Al_2O_3$ using Raman spectroscopy. Average crystallite size and the strain are also obtained from XRD. Both confined acoustic and optical phonons of Si are investigated. The sizes estimated from XRD, confined acoustic phonon and from fitting of confined optical phonon line shape are compared. Stress is estimated from the blue-shift of the Raman peak. The complex frequency model (CFM) is used for calculating the confined acoustic phonon frequencies. The possible reason for the dependence of the stress on the nanocrystal size is also discussed.

## II. EXPERIMENTAL DETAILS

1 MeV $Si^{+2}$ ions were implanted in (0001) oriented $c-Al_2O_3$ matrix for fluences of $5\times10^{16}$ (sample I) and $1\times10^{17}$ (sample II) $cm^{-2}$ at ambient temperature using Tandetron Accelerator (High Voltage Engineering Europa, The Netherlands). The implanted samples were annealed at 1373 K, for 3 hrs in $Ar-H_2$ mixture (8% $H_2$) gas. X-ray diffraction (XRD) measurements were carried out using MAC Science MXP18 X-ray diffractometer with Cu-K$\alpha$ radiation from a rotating anode X-ray generator. The nanocrystals were also imaged using a transmission electron microscope (JOEL 2000FX).

The low-frequency Raman scattering (LFR) measurement were carried out in a back scattering geometry using a Spex 14018 double monochromator equipped with a Argon ion laser (488 nm) and photomultiplier tube (Hamamatsu R943-02) detector. Raman scattering measurements were performed using 532 nm line of solid-state laser as the excitation and analyzed using a double-monochromator (Jobin-Yvon U1000), equipped with liquid nitrogen cooled CCD detector.

## III. RESULT AND DISCUSSION

In as-implanted samples Si atoms remains uniformly distributed at the implantation depth. Upon annealing Si atoms diffuse in $Al_2O_3$ hosts and form isolated nanoclusters. In nanocrystals, low frequency vibrational modes are associated with collective motion of atoms. These are acoustic in nature and are known as confined acoustic phonons. Lamb analyzed the low frequency vibrational mode of a homogenous elastic spherical particle and showed that the frequencies of these modes are determined by the eigen values $\eta_{l,n}$ of an eigen-value equation and depend inversely on the particle diameter.[13] Subsequently Duval showed that for spherical nanoparticles only the $l = 0$ and $l = 2$ spheroidal modes are Raman active.[14] Recently, a complex frequency model (CFM) has been developed to understand the effect of host matrix on the vibrational frequencies of nanocrystals.[15]

Figure 1 shows the LFR spectra of samples I and II respectively. Two distinct broad confined acoustic phonon peaks are found for each sample. Using a least square fitting program and taking an exponential background, peak positions were obtained. For sample I the peaks appears at 31.1±0.3 and 22.5±0.5 $cm^{-1}$ and for sample II they occur at 20.3±0.35 and 15.5±0.3 $cm^{-1}$. The higher mode frequencies are assigned to $l = 0$ breathing mode whereas the lower frequency modes are assigned to $l = 2$ quadrupolar mode. CFM was used for estimating the particle sizes corresponding to the observed mode frequencies. The calculation shows that the mode frequencies for sample I corresponds to a size of 4.0±0.8 nm whereas those for sample II are consistent with a size of 6.0±0.6 nm. As the sample II has higher dose than sample I, larger particle sizes are expected in sample II.

Figure 2 shows the Raman spectra arising from $F_{2g}$ optical phonon of Si nanocrystals in samples I and II. For the sake of comparison Raman spectra of bulk Si, recorded using the same setup, is also included in the figure. Distinct peaks are observed at 525.9±0.1 and 524.0±0.1 cm$^{-1}$ for samples I and II respectively. On the other hand, the Raman peak for bulk-Si appears at 520 cm$^{-1}$. One can also see that Raman lineshape becomes broader for smaller particle size. In order to analyze the changes in the Raman lineshape one needs to obtain the modes allowed in the finite system. Using the boundary condition that the phonon amplitude is zero at particle boundary, the discrete allowed wavevectors in the Brillouin zone are obtained. In contrast to bulk crystals where only $q = 0$ (zone center) optical phonon are allowed, in the nanocrystals several phonon mode frequencies corresponding to the discrete allowed wavevectors contribute to the Raman intensity. This results in asymmetric broadening of the Raman lineshape. One can in principle calculate the Raman lineshape by adding the contribution of each of these modes (having their own natural linewidth) with suitable $q$-dependent weight or intensity factors. However, the intensity of the individual modes is not straightforward to calculate and hence often a phenomenological approach based on spatial correlation model has been used to analyze the Raman lineshape.[16] Here the phonon amplitude is taken as a suitable decaying function (confinement function) and contribution from all the phonons away from the zone center are integrated over the Brillouin zone to yield the Raman lineshape. This alternative formalism also gives a reasonable description of the phonon lineshape.[16-18] Following the formalism of Campbell and Fauchet[18] one can use a Gaussian confinement function, for a spherical nanocrystal of diameter $d$ as

$$W(r) = \exp\left(-\alpha r^2 / d^2\right), \tag{1}$$

where the value of $\alpha$ decides how rapidly the wave function decays as one approaches the boundary. A large number of results have been satisfactorily explained on the basis of Gaussian confinement model using $\alpha = 8\pi^2$. For one-phonon Raman scattering the $q$-dependent weight factor for estimating the contribution of phonons away from zone center is the Fourier transformation of the confinement function. For Gaussian confinement function, the weight factor is

$$|C(q)|^2 = \exp\left(-q^2 d^2 / 2\alpha\right) \qquad (2)$$

The first-order Raman spectrum is calculated by integrating these contributions over the complete Brillouin zone, as

$$I(\omega) = \int \frac{|C(q)|^2}{[\omega - \omega(q)]^2 + (\Gamma_0 / 2)^2} d^3q \qquad (3)$$

where $\omega(q)$ is the phonon dispersion curve and $\Gamma_0$ is the natural line width of zone center optical phonon in bulk Si.

The expected Raman spectra as calculated using phonon confinement model are also shown in Fig. 2. as dashed curves. Note that except for the peak position, the calculated phonon line shapes for 4 and 6 nm particles match well with the observed Raman line shapes of samples I and II respectively. Thus the particle sizes estimated from phonon confinement model agree with those obtained from CFM.

We now examine the difference between the peak centers of the calculated and observed spectra. In contrast to the phonon confinement model, which predicts a red shift of optical phonon as a function of size of nanoparticle, a blue shift of optical phonon is observed for both the samples. The peak positions expected from phonon confinement model for 4 and 6 nm Si nanoparticles are 519 and 519.4 cm$^{-1}$ respectively whereas the observed phonon peaks are found at 525.9 and 524 cm$^{-1}$ for samples I and II respectively. This blue shift of the peak-centers of optical phonon can arise if the embedded Si-nanocrystals experience compressive stress. The pressure exerted by the Al$_2$O$_3$ matrix on nc-Si can be obtained from the blue shift of optical phonon frequency. The pressure induced shift of phonon frequency in Si is given as[19]

$$\omega = 520 + 5.5\, P, \qquad (4)$$

where $\omega$ is in cm$^{-1}$ and $P$ in GPa. For sample I (4 nm) and sample II (6 nm) the pressures are calculated to be 1.2 and 0.8 GPa respectively. The full curves in Fig. 2 are the confined phonon line shapes for the stressed Si-nanoclusters of size 4 and 6 nm. One can see a good agreement between the calculated and experimental spectra.

Figure 3 shows the diffraction patterns of the samples I and II for (111) reflection of Si. The broadening of the XRD peak also has contribution from other factors such as microstrain. Its contribution can be estimated if widths of several diffraction peaks are considered. However, in the present case other peaks were found to be absent suggesting growth of (111) oriented Si clusters in (0001) oriented single crystal Al$_2$O$_3$ host. In order to account for the asymmetry of the diffraction line, the contribution of quasi-line[20] was considered in the diffraction profile analysis. The particle sizes, estimated from Scherrer's formula are found to be 6.4 and 7.9 nm for samples I and II respectively. Although XRD gives some what larger sizes, the trend remains the same as that found from LFR and phonon confinement model. In order to get an idea about actual size, samples were also examined in TEM. Fig. 4 shows the TEM micrographs of sample I. The average particle sizes and their standard deviation obtained from histogram were 4±1.5 and 7±1 nm for samples I and II respectively. In addition to the broadening the diffraction pattern also exhibited a small shift (~0.11 degree) to larger 2θ with respect to reported JCPDS (Card No. 27-1402) positions. This shift also suggests a compressive strain on the nanoparticles. The pressure estimated from the shift of the diffraction peaks and the reported Bulk modulus (100 GPa) turn out to be ~1.0±0.2 GPa for both the samples consistent with that estimated from phonon blue-shift.

The earlier studies on Si-nanocrystals embedded in $SiO_2$ nanowires[9] were on a system where both the core (particle) and the shell (host) thicknesses varied leading to the ambiguity about the origin of stress. On the other hand, the present results on Si-nanocrystals buried deep inside the $Al_2O_3$ matrix clearly show that the stress essentially varies because of the size. The stress on the embedded nanoparticles arise because these form during annealing at elevated temperature and compressive stress develops due to the differences in the elastic constant and the thermal expansion coefficients of the particles and the host during cooling to ambient temperature. Furthermore large particles are expected to have less stress as compared to small particles, because during the growth of larger particles the stress gets relieved by deforming the surrounding host.

## IV. CONCLUSION

To summarize, we have investigated Si-nanoparticles embedded in c-$Al_2O_3$ host by examining the confined acoustic and $F_{2g}$ optical phonon using Raman spectroscopy. From a detailed analysis of Raman spectra of the $F_{2g}$ phonon, the effects of phonon confinement and compressive stress on the Si-nanoparticles are separated and quantified.

The phonon confinement results in the red-shift of the peak-centre and asymmetric broadening whereas the compressive stress leads to a blue shift of the Raman spectra.

## ACKNOWLEDGEMENTS

Authors thank Dr. C. S. Sundar and Dr. T. Jayakumar for interest in the work, Dr. P. R. Vasudeva Rao for support and Dr. Baldev Raj for encouragement. Help of C. P. Chen, Centre for Condensed Matter Science, Taipei, in obtaining the TEM micrographs is also acknowledged.

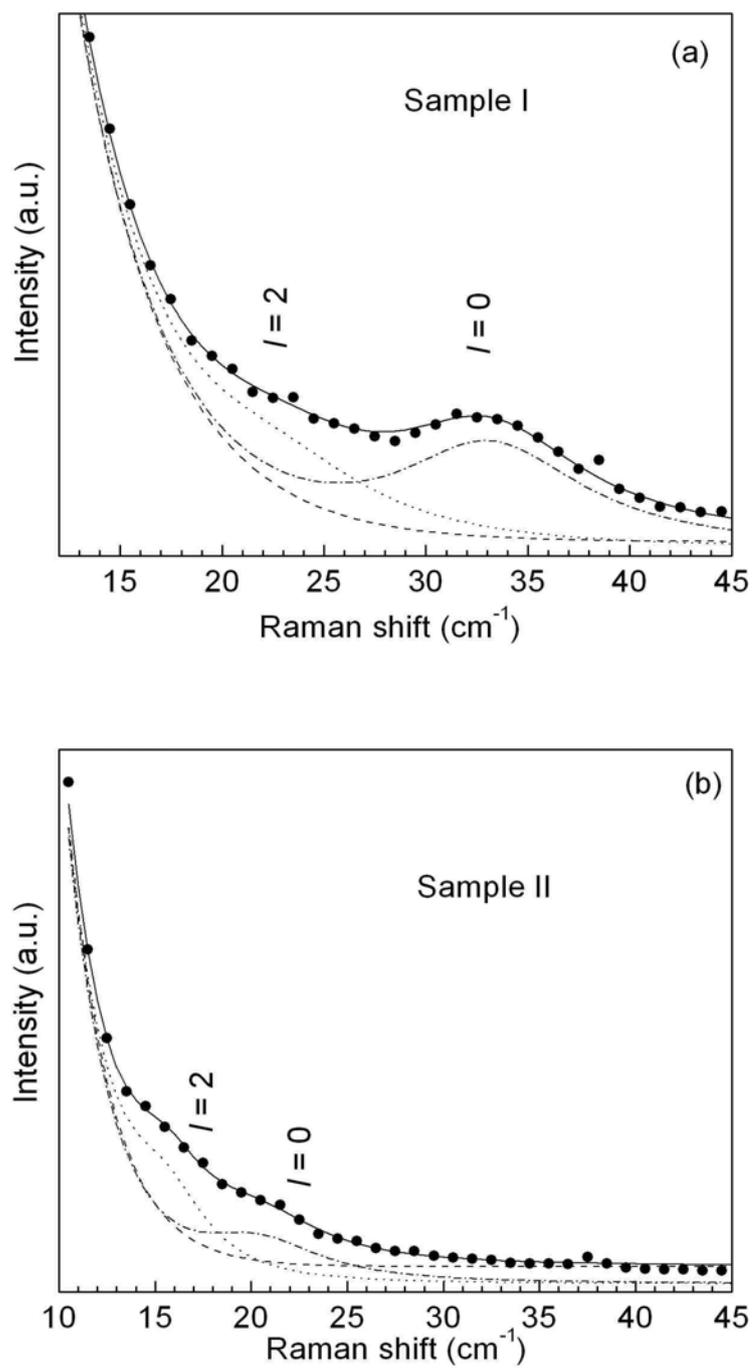

Fig. 1. Low-frequency Raman spectra of Si nanoclusters embedded in c-$Al_2O_3$. (a): sample I and (b): sample II. Filled symbols are the data. The dashed and dotted curves are the exponential background and the fitted peaks. Full curves are the total fitted spectra.

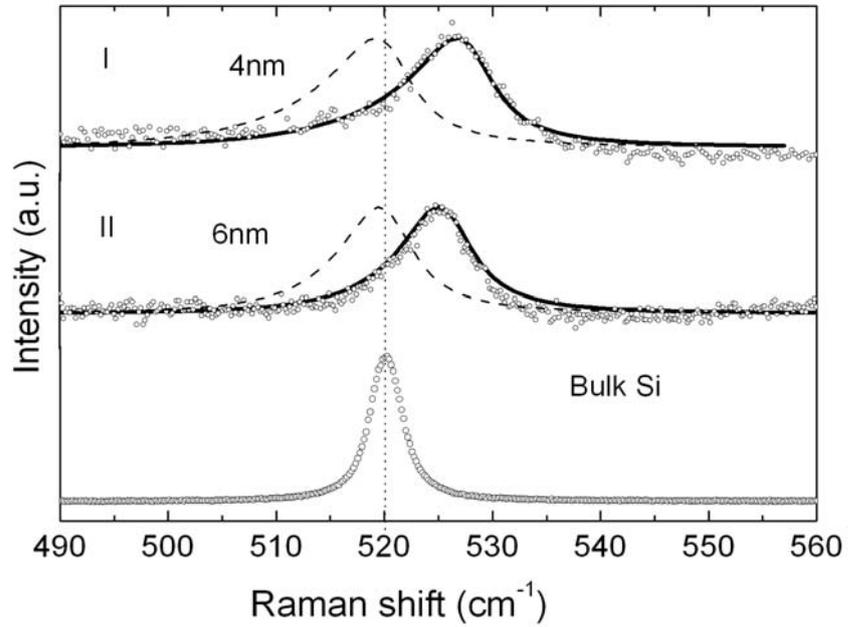

Fig. 2. $F_{2g}$ Optical phonon Raman spectra for samples I, II and bulk-Si. Open symbols represents experimental data. Dashed curves are confined optical phonon line shapes for 4 and 6 nm nc-Si without taking stress into account. The full curves are the calculated profiles for stressed nc-Si.

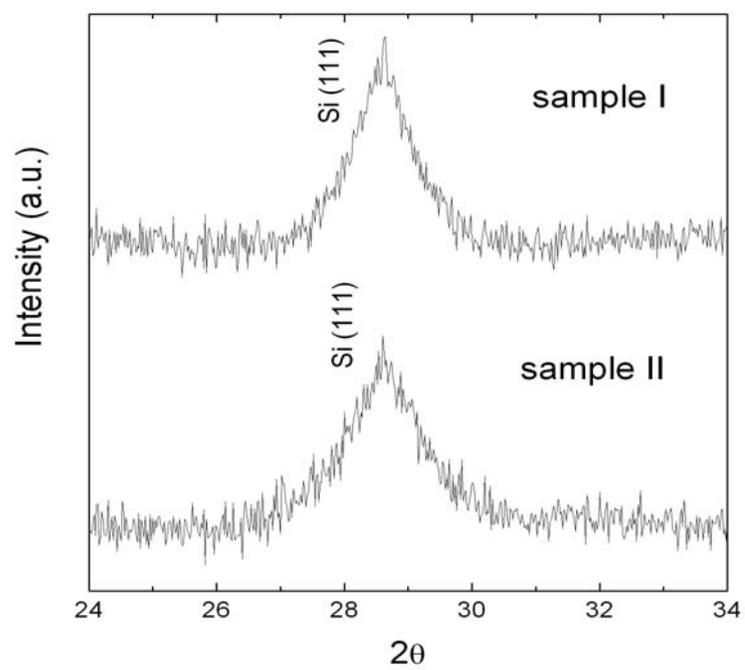

Fig. 3. X-ray diffraction patterns of samples I and II in the region of (111) reflection of silicon.

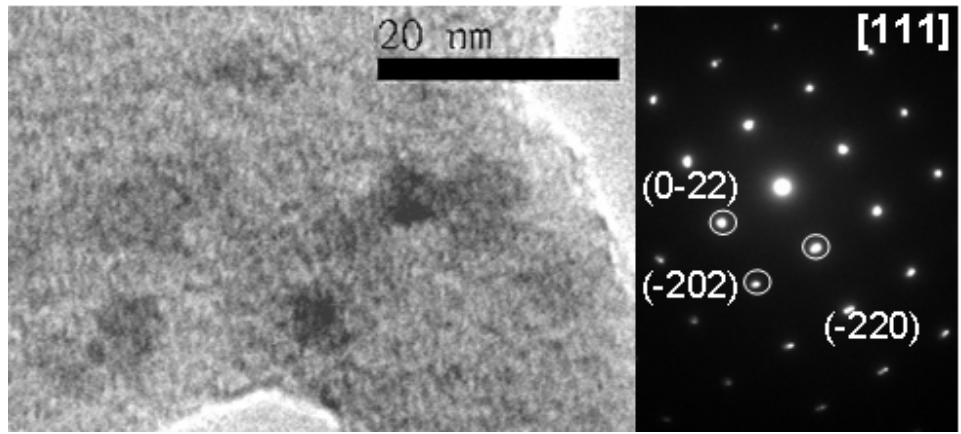

Fig. 4. TEM image of sample I and the selected area diffraction pattern from the Si-